\documentstyle[twoside,fleqn,espcrc2,graphicx]{article}

\def\la{\mathrel{\mathpalette\fun <}}
\def\ga{\mathrel{\mathpalette\fun >}}
\def\fun#1#2{\lower3.6pt\vbox{\baselineskip0pt\lineskip.9pt
  \ialign{$\mathsurround=0pt#1\hfil##\hfil$\crcr#2\crcr\sim\crcr}}}

\newcommand{\AmS}{{\protect\the\textfont2
  A\kern-.1667em\lower.5ex\hbox{M}\kern-.125emS}}

\title{Cosmic rays and neutrino interactions beyond the standard model}

\author{G\"unter Sigl\address{
D\'epartement d'Astrophysique Relativiste et de Cosmologie, CNRS\\
Observatoire de Paris, 92195 Meudon Cedex, France}}
       
\begin{document}

\begin{abstract}
Some solutions of the hierarchy problem of particle physics
can lead to significantly increased neutrino cross
sections beyond the electroweak scale. We discuss some
consequences for and constraints resulting from cosmic ray physics.
\end{abstract}

\maketitle

\section{Introduction}
It has been suggested that the neutrino-nucleon
cross section could be enhanced by new physics beyond the
electroweak scale in the center of mass frame, or above about
a PeV in the nucleon rest frame. A specific implementation
of this possibility is given in theories with $n$ additional
dimensions and a quantum gravity scale $M\sim\,$TeV
that has recently received much attention in the literature~\cite{tev-qg}
because it provides an alternative solution (i.e., without 
supersymmetry) to the hierarchy problem
in grand unifications of gauge interactions. 
In such scenarios, the exchange of bulk gravitons (Kaluza-Klein
modes) can lead to an extra contribution to any two-particle cross section
given by~\cite{ns}
\begin{equation}
  \sigma_g\simeq\frac{4\pi s}{M^4}\simeq
  10^{-27}\left(\frac{{\rm TeV}}{M}\right)^4
  \left(\frac{E}{10^{20}{\rm eV}}\right)\,{\rm cm}^2\,,
  \label{sigma_graviton}
\end{equation}
where the last expression applies to a neutrino
of energy $E$ hitting a nucleon at rest. Note that a neutrino
would typically start to interact in the atmosphere
and therefore become a primary candidate for the
highest energy cosmic rays
for $\sigma_{\nu N}\ga10^{-27}\,{\rm cm}^2$, i.e. for
$E\ga10^{20}\,$eV, assuming $M\simeq1\,$TeV.

The total charged-current neutrino-nucleon cross section is given by
the sum of Eq.~(\ref{sigma_graviton}) and the cross section
within the Standard Model, which can be estimated by~\cite{gqrs}
\begin{equation}
  \sigma^{SM}_{\nu N}(E)\simeq2.36\times10^{-32}
  \left(\frac{E}{10^{19}\,{\rm eV}}\right)^{0.363}
  \,{\rm cm}^2\label{cccross}
\end{equation}
in the energy range $10^{16}\,{\rm eV}\la E\la10^{21}\,$eV.

The total cross section is dominated
by a contribution of the form Eq.~(\ref{sigma_graviton})
at energies $E\ga E_{\rm th}$, where, for $M\ga1\,$TeV,
the threshold energy can be approximated by
\begin{equation}
  E_{\rm th}\simeq2\times10^{13}\,
  \left(\frac{M}{{\rm TeV}}\right)^{6.28}\,
  {\rm eV}\,.\label{Eth}
\end{equation}
This would be reflected by a linear energy dependence
of the typical column depth of induced shower development
if the optical depth in the detection medium is of
order unity, or by a flattening of the differential
detection rate by one power of the energy if the
optical depth is smaller than unity. Comparison with
observations would either reveal signatures for these
scenarios or constrain them in a way complementary to
and independent of many studies
on signatures in human made accelerators~\cite{coll-extra-dim}
or other laboratory experiments that have recently appeared in the
literature.

\section{A Bound from the ``Cosmogenic'' Neutrino Flux}

\begin{figure}[t]
\includegraphics[width=7.5cm]{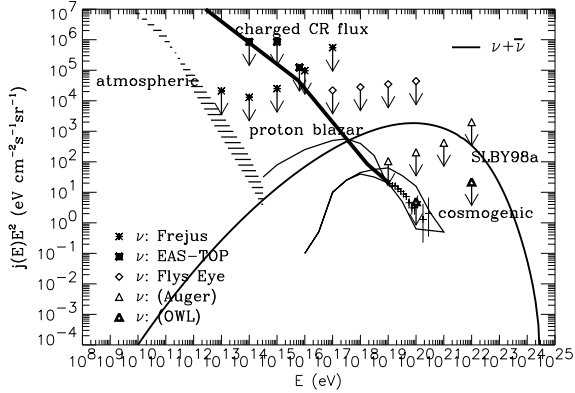}
\caption[...]{Predictions for the differential fluxes summed over all
neutrino flavors (solid lines) from the sources discussed in the text.
1 sigma error bars are the combined data from the Haverah Park~\cite{haverah},
the Fly's Eye~\cite{fe}, and the AGASA~\cite{agasa} experiments
above $10^{19}\,$eV. Also shown are piecewise power law fits to the observed
charged CR flux (thick solid line).
Points with arrows represent approximate upper limits on the
diffuse neutrino flux from the Frejus~\cite{frejus}, the
EAS-TOP~\cite{eastop2}, and the Fly's
Eye~\cite{baltrusaitis} experiments, as indicated. The projected
sensitivity for the Pierre Auger project is using the acceptance
estimated in Ref.~\cite{auger-neut}, and the one for the OWL concept
study is based on Ref.~\cite{owl}, both assuming observations
over a few years period.
\label{fig:fig1}}
\end{figure}

Fig.~\ref{fig:fig1} shows neutrino fluxes for the atmospheric
background at different zenith angles~\cite{lipari} (hatched
region marked ``atmospheric''), for proton blazars that are
photon optically thick to nucleons~\cite{protheroe2} and whose
flux was normalized
to recent estimates of the blazar contribution to the diffuse
$\gamma-$ray background~\cite{muk-chiang}
(``proton blazar''), for neutrinos created as secondaries from the decay
of charged pions produced by ultra-high energy (UHE) nucleons interacting
with the cosmic microwave background~\cite{pj}
(``cosmogenic''), and for a model where UHE cosmic rays are produced
by decay of particles close to the Grand Unification Scale
(``SLBY98'', see Ref.~\cite{slby} for details).

Apart from the atmospheric neutrino flux only the cosmogenic
neutrinos are guaranteed to exist due to the known existence
of UHE cosmic rays, at least if these contain nucleons and are not
exclusively of galactic origin.

The non-observation of deeply penetrating air showers by the
experiments indicated in Fig.~\ref{fig:fig1} in the presence
of this cosmofenic flux can now be translated into an upper limit on
the total neutrino-nucleon cross section $\sigma_{\nu N}\equiv
\sigma^{SM}_{\nu N}+\sigma_g$ by scaling the
diffuse neutrino flux limits from the
Standard Model cross section Eq.~(\ref{cccross}). Using the
conservative, lower estimate of the cosmogenic flux
in Fig.~\ref{fig:fig1} yields
\begin{equation}
  \sigma_{\nu N}(E=10^{19}\,{\rm eV})\la2.4\times10^{-29}
  \,{\rm cm}^2\,,\label{crosslim1}
\end{equation}
as long as $\sigma_{\nu N}(E=10^{19}\,{\rm eV})\la10^{-27}\,{\rm cm}^2$,
such that neutrinos would give rise to deeply penetrating
air showers. Using Eq.~(\ref{sigma_graviton}) results in
\begin{equation}
  M\ga1.4\,{\rm TeV}\,.\label{Mlim1}
\end{equation}

It is interesting to note that these limits do not depend
on the number $n$ of extra dimensions, in contrast to
some other astrophysical limits such as from graviton emission
from a hot supernova core into the extra dimensions
which depend more explicitly on phase space integrations
(see Sect.~3 below).

As can be seen from Fig.~\ref{fig:fig1}, with an experiment such
as OWL, the upper limit on the cross section Eq.~(\ref{crosslim1})
could improve by about 4 orders of magnitude, and the lower
limit on $M$ consequently by about a factor 10.

\section{Comparison with Other Astrophysical and Laboratory Bounds}
There are also astrophysical constraints on $M$ which result
from limiting the emission of bulk gravitons into the extra dimensions. 
The strongest constraints in this regard come from nucleon-nucleon
bremsstrahlung
in type II supernovae~\cite{astro-extra-dim}. These contraints read
$M\ga50\,$TeV, $M\ga4\,$TeV, and
$M\ga1\,$TeV, for $n=2,3,4$, respectively, and,
therefore, $n\geq4$ is required if neutrino primaries
are to serve as a primary candidate for the UHE cosmic ray events observed
above $10^{20}\,$eV (note that $n=7$ for the superstring and
$n=22$ for the heterotic string). This assumes that all extra
dimensions have the same size given by
\begin{eqnarray}
  r&\simeq&M^{-1}\left(\frac{M_{\rm Pl}}{M}\right)^{2/n}
  \nonumber\\
  &\simeq&2\times10^{-17}\left(\frac{{\rm TeV}}{M}\right)
  \left(\frac{M_{\rm Pl}}{M}\right)^{2/n}{\rm cm}
  \,,\label{rextra}
\end{eqnarray}
where $M_{\rm Pl}$ denotes the Planck mass. 
The above lower bounds on $M$ thus translate into the corresponding
upper bounds $r\la3\times10^{-4}\,$mm,
$r\la4\times10^{-7}\,$mm, and $r\la2\times10^{-8}\,$mm,
respectively.

UHE cosmic rays and neutrinos together with other astrophysical
and cosmological constraints thus provide an interesting testing
ground for theories involving extra dimensions which represent
one possible kind of physics beyond the Standard Model. 
In this context, we mention that in theories with large compact extra
dimensions mentioned above, Newton's law of gravity is expected to be
modified at distances smaller than the length scale given by
Eq.~(\ref{rextra}). Indeed, there are laboratory 
experiments measuring gravitational interaction at small
distances (for a recent review of such experiments see
Ref.~\cite{lcp}), which also probe these theories. Thus, future UHE
cosmic ray experiments and gravitational experiments in the
laboratory together 
have the potential of providing rather strong tests of these theories. 
These tests would be complementary to constraints
from collider experiments~\cite{coll-extra-dim}.

\end{document}